\documentclass[twocolumn,preprintnumbers,amsmath,amssymb]{revtex4}
\input{epsf}

\newcommand{\be}{\begin{equation}}
\newcommand{\ee}{\end{equation}}
\newcommand{\bea}{\begin{eqnarray}}
\newcommand{\eea}{\end{eqnarray}}
\newcommand{\bef}{\begin{figure}}
\newcommand{\eef}{\end{figure}}

\newcommand{\simge}{\,{}^>_{\sim}\,}
\newcommand{\simle}{\,{}^<_{\sim}\,}

\def\h#1{$^{#1}$H}
\def\he#1{$^{#1}$He}
\def\li#1{$^{#1}$Li}

\def\eps@scaling{0.96}

\def\showone#1{
  \centering
  \leavevmode
  \epsfxsize=\eps@scaling\linewidth
  \epsfbox{#1.eps}
}

\def\epstwo@scaling{0.48}

\def\showtwo#1#2{
  \centering
  \leavevmode
  \epsfxsize=\epstwo@scaling\linewidth
  \epsfbox{#1.eps} \hfil
  \epsfxsize=\epstwo@scaling\linewidth
  \epsfbox{#2.eps}
}

\begin{document}

\title{Neutralinos, Big Bang Nucleosynthesis and \li6
in Low-Metallicity Stars}
\author{Karsten Jedamzik} 
\affiliation{Laboratoire de Physique Math\'emathique et Th\'eorique, C.N.R.S.,
Universit\'e de Montpellier II, 34095 Montpellier Cedex 5, France}

\begin{abstract}
The synthesis of \li6 during the epoch of
Big Bang nucleosynthesis (BBN) due to residual annihilation of dark
matter particles is considered. By comparing the predicted \li6
to observations of this
isotope in low-metallicity stars, generic constraints on s-wave dark matter
annihilation rates into quarks, gauge bosons, and Higgs bosons are derived. 
It may be shown that, for example,
wino dark matter in anomaly-mediated SUSY breaking scenarios with masses 
$m_{\chi}\simle 250\,$GeV or
light neutralinos with $m_{\chi}\simle 20\,$GeV annihilating
into light quarks are, taken face value, ruled out. These constraints
may only be circumvented if 
significant \li6 depletion has occurred
in all three 
low-metallicity stars in which this isotope has been observed to date.
In general, scenarios invoking non-thermally generated neutralinos with
enhanced annihilation rates for a putative
explanation of cosmic ray positron or 
galactic center as well as diffuse background gamma-ray signals by present-day
neutralino annihilation will have to
face a stringent \li6 overproduction problem. On the other hand, it is
possible that \li6 as observed in low-metallicity stars is entirely due
to residual dark matter annihilation during BBN, even for 
neutralinos undergoing a standard thermal
freeze-out.
\end{abstract}


\maketitle

The nature of the ubiquitous cosmological
dark matter is one of the outstanding questions 
in cosmology. Though there exist a multitude of proposed 
candidates, much consideration has been given to the supersymmetric 
(SUSY) neutralino,
provided it is the lightest supersymmetric particle (LSP).
SUSY extensions
of the standard model are particularly
successful in overcoming a number of shortfalls of the standard model,
such as the hierarchy problem and grand unification.
Neutralinos are appealing since neutral, likely stable, 
and endowed with annihilation/scattering
cross sections which (a) makes it likely have the right cosmological
abundance and (b) makes it be detectable in the not-to-distant future
by either direct- (e.g. scattering in cryogenic detectors) 
or indirect- (e.g. observation of positrons or $\gamma$-rays due
to residual neutralino annihilation in the Galaxy) 
detection means~\cite{darkrev}.

In general, the post-freeze-out number-to-radiation entropy
$n_{\chi}/s$ of a stable and initially abundant particle subject 
to self-annihilation
is given by
\begin{equation}
Y_{\chi}^f = (n+1)\biggl({H\over \langle\sigma v\rangle s}\biggr)_f\, .
\label{eq:yf}
\end{equation}
where $\langle\sigma v\rangle$ denotes it's thermally averaged annihilation
rate, $H$ the Hubble expansion rate at freeze-out, $n=0(1)$ for
s-wave (p-wave) annihilation, i.e. 
$\langle\sigma v\rangle =\sigma_0x^{-n}$ with $x=m_{\chi}/T$, respectively,
and it is understood that quantities are evaluated at the moment of
freeze-out (f) itself. Though Eq.~(\ref{eq:yf}) could have 
been obtained by simply equating a typical annihilation rate 
$1/Y_{\chi}dY_{\chi}/{\rm d}t$ with the Hubble rate it turns out to be exact.  
The abundance of Eq.~(\ref{eq:yf})
yields a $\chi$ contribution to the present 
critical density $\Omega_{\chi}$ of
\begin{equation}
\Omega_{\chi}h^2 = 9.95\times 10^{-3}(n+1)x_f^{(n+1)}{\sqrt{g_H^f}\over g_S^f}
\langle\sigma v\rangle_{-25}^{-1}
\label{Omega}
\end{equation}
where $x_f = m_{\chi}/T$ at freeze-out, 
$\langle\sigma v\rangle_{-25}$ is the annihilation rate in units of
$10^{-25}{\rm cm^3/s}$ at $T=m_{\chi}$
and $g_H^f$, $g_S^f$ are energy- and
entropy- radiation statistical weights at freeze-out, respectively.

Considering freeze-out from annihilations with thermal equilibrium 
initial conditions, a particle in the hundreds of GeV range with 
$\langle\sigma v\rangle_{-25}\approx 0.2$ freezing
out at $x_f\approx 20$ when $g_H^f\approx g_S^f\approx 86.25$ due
to the known standard model degrees of freedom comes close to the
recently by WMAP determined matter density of 
$\Omega_mh^2\approx 0.1126^{+0.0161}_{-0.0181}$~\cite{Sper:03}
(with h the Hubble
constant in units of $100\,{\rm km\,s^{-1}Mpc^{-1}}$ and excluding
baryons). In the case
of mSUGRA models, 
where the LSP is typically
the bino, this may
be the case in thin 
strips of parameter space when 
$m_{\chi}\simle 500\,$GeV
though 
the bino $\langle\sigma v\rangle_{-25}$ is typically smaller than
0.2~\cite{darkrev}. Other parameter space with viable neutralino abundances
$\Omega_mh^2\simle 0.113$ (corresponding to 
$\langle\sigma v\rangle_{-25}\simge 0.2$) in mSUGRA exist in the so-called
FOCUS point area at several TeV unifying scalar masses $m_0$, with the
LSP having a large Higgsino component.
Leaving the fairly restrictive limitation of mSUGRA a wider array of 
possibilities may occur. When gaugino mass unification at the GUT scale
is dropped~\cite{Cerdeno:2004zj}, 
the LSP is frequently the wino or higgsino, albeit often
with an s-wave annihilation rate of the order 
$\langle\sigma v\rangle_{-25}\sim 1-100$. Similar large annihilation
rates may occur in anomaly-mediated SUSY breaking schemes 
(AMSB)~\cite{Gherghetta:1999sw,Moroi:1999zb,Ullio01} 
where
the LSP is likely the wino, or even in mSUGRA when annihilation may
occur on Higgs poles (e.g. the so-called A-funnel region for annihilation
on the CP-odd Higgs)
for $m_{\chi}\approx m_H/2$~\cite{poles}. 
It is also possible that the neutralino LSP is very light~\cite{Bott:02}. 

Many of these
cases have in common that for $m_{\chi}$ not too high, thus avoiding
fine-tuning for the electroweak symmetry breaking to occur,
annihilation rates are too large to 
produce $\Omega_mh^2\approx 0.113$ during thermal freeze-out.
Nevertheless, there is a number of proposed scenarios generating
neutralino dark matter either by non-thermal means or due to the
existence of extra degrees of freedom in the early Universe.
Q-balls~\cite{Qballs}, 
for example, non-topological solitons which occur in the
spectrum of SUSY theories, may be easily formed after an inflationary epoch
of the Universe. As they are made of squarks, and are unstable in
SUGRA and AMSB, their R-parity and baryon-number conserving evaporation
at late times $T_D\simle 1\,$GeV may not only generate many LSP's but
also the observed cosmic baryon asymmetry. The thus generated LSP's
may undergo further self-annihilation to reach an asymptotic final
abundance dependent on $T_D$~\cite{Qball_neutralino}. 
Similarly, in AMSB scenarios, 
gravitinos are typically heavy $m_{\tilde{G}}\simge 50\,$TeV, such that
their decay occurs before BBN. In such scenarios, which circumvent
the gravitino problem, neutralinos are produced in $\tilde{G}$
(and/or moduli) decays well after the conventional thermal
freeze-out of neutralinos from 
equilibrium~\cite{Gherghetta:1999sw,Moroi:1999zb}. 
When additional substantial
non-thermal injection of neutralinos at $T_D$ occurs, their final
present day abundance may still be calculated via Eq.~(\ref{eq:yf}) and
Eq.~(\ref{Omega}) with $x_f$ given by
\begin{equation}
x_f ={\rm max}[x_D,x_{tf}]
\end{equation}
where $x_D = m_{\chi}/T_D$, and $x_{tf}$ quantifies the appropriate 
freeze-out temperature during thermal
freeze-out~\cite{xtf},
and where it has been implicitly assumed that the 
pre-annihilation neutralino abundance exceeds that of Eq.~(\ref{eq:yf}) and
Eq.~(\ref{Omega}). LSP abundance estimates may also be changed (i.e. increased) when
compared to the standard evolution of the Universe, when additional degrees
of freedom are present in the early Universe. This may be, for
example, due to the existence of a quintessence field in the stage of
kination, 
or quite generally, due to extra
degrees of freedom $g_{H}$ (with $g_{S}$ unchanged)
contributing to the Hubble expansion~\cite{quintessence}. 
There are currently
no limits on $g_{H}$ at high $T$ other
than those from the (much later) epoch at BBN $T\simle 1\,$MeV. 

Though the post-freeze-out $Y_{\chi}^f$ stays essentially unchanged,
residual $\chi$ annihilations occur up to the present day and may
have considerable impact on the process of Big Bang nucleosynthesis (BBN). The
effects of residual annihilation of $\chi$'s
during BBN have been considered some time 
ago~\cite{Reno:88,Frieman:1989fx}. Nevertheless, whereas
earlier studies were based on the isotopes \h2 and \he4 , it has been
pointed out that \li6 production during electromagnetic~\cite{Dimo:88,Jeda:00} 
or hadronic~\cite{Dimo:88,Khlo:94,Jeda:04,Kawasaki:2004yh}
energy injection during BBN is particularly efficient~\cite{remark7}.
The isotope of \li6 may be produced via non-thermal reactions
(subject to energy threshold), such as \h3$(\alpha ,n)$\li6
or \he3$(\alpha ,p)$\li6. Energetic \h3 and \he3 are readily produced
via nuclear \he4 spallation by energetic nucleons
or \he4 photodisintegration. Since the
photodisintegration of \he4 is only efficient for 
$T\simle 0.3\,$keV particular importance in the annihilation case
is due to hadronic spallation. This is because residual annihilation
is stronger at earlier times. Resulting \li6 abundances are given 
formally by
a time integral
\begin{equation}
{n_{\rm ^6Li}\over n_{\rm H}} 
= \int {\rm d}t\biggl({{\rm d\,Ann}\over{\rm d}t}\biggr)
                          \biggl({n_{p,n}\over {\rm Ann}}\biggr)
                          \biggl({n_3\over n_{p,n}} \biggr)
                          \biggl({n_6\over n_3}\biggr)
\label{li1}
\end{equation}
where the factors on the right-hand-side from left to right, denote
(a) the $\chi$ annihilation rate
\begin{equation}
\frac{{\rm d\, Ann}}{{\rm d}t} = \frac{\langle\sigma v\rangle}{2}
\biggl(\frac{\rho_{\chi}}{m_{\chi}}\biggr)^2
\label{ann}
\end{equation}
with $\rho_{\chi}$ neutralino density, (b) the generated energetic protons
$p$ and neutrons $n$ per annihilation, (c) the produced energetic mass-three nuclei
per generated energetic nucleon, and (d) the produced \li6 per energetic
mass-three nucleus, respectively. Here the first factor is determined by the 
annihilation rate under the assumption that $\chi$ is the dark matter
whereas the second factor may be obtained assuming a 
particular $\chi$ annihilation
channel and computing the energetic nucleon spectrum via a
hadronic flux tube Monte Carlo code  such as PYTHIA~\cite{pythia}. 
Concerning the third
and fourth factors I have recently presented first~\cite{Jeda:04} 
results and a first 
description of a newly developed Monte-Carlo code describing the cascading of energetic
nucleons and nuclei on background thermal nucleons, nuclei, and
electromagnetically interacting particles which allows me to evaluate
Eq.~(\ref{li1}). This code includes, Coulomb and Thomson stopping of
fast charged nucleons and nuclei, elastic- and inelastic- nucleon-nucleon
scattering and elastic- and break-up- processes in nucleon-\he4 scattering.
Of all mass three nuclei produced a small fraction react on \he4 to form
\li6. The probability to do so, may be approximated by
\bea
P_{\rm ^6Li} = \biggl({n_6\over n_3}\biggr) =
\int{\rm d}E_i {{\rm d} n_3\over{\rm d}E_i} 
\int_{E_{th}}^{E_i}{\rm d}E{1\over l_{\rm ^6Li}\,
{{\rm (d}E/{\rm d}x)_C}} \nonumber\\
\times\, 
{\rm exp}\biggl(-\int_E^{E_i}{\rm d}E^{\prime}{1\over l_{nuc}\,
({\rm d}E/{\rm d}x)_C}\biggr)\, \label{Prob}
\eea
a convolution over the initial energy distribution of mass-three nuclei, 
${\rm d} n_3/{\rm d}E_i$, - the probability that a mass-three nucleus during it's
passage through energy space due to the dominant
multiple Coulomb interactions
with energy loss per unit length ${\rm d}E/{\rm d}x$ undergoes
a reaction to form \li6 (where 
$l_{\rm ^6Li} = 1/(\sigma_{\rm ^6Li} n_{\rm ^4He })$
is the mean free path towards formation of \li6) - while, the exponential
factor, not having undergone
already another nuclear interaction
(mean free path due to all
nuclear interactions is $l_{nuc}$)~\cite{remark1}.
Compared to the Coulomb losses other nuclear interactions in Eq.~(\ref{Prob})
are typically not very important, except for charge exchange between
the mirror nuclei
\h3$(p,n)$\he3 and to a lesser degree elastic p-\h3 and p-\he3 scattering.
in only a narrow temperature range between
several keV and $\approx 20\,$keV. Nevertheless, 
though the freshly synthesized
energetic \li6 typically survives almost completely 
$p$-spallation~\cite{remark3}
it may only survive \li6$(p,\alpha )$\he3 thermal reactions (after it's
rapid thermalization)
when $T\simle 10\,$keV. It has been 
realized~\cite{Reno:88,Kawasaki:2004yh} that in the 
above-given temperature window 
Coulomb losses loose some of their efficiency (by roughly a factor
3-10) for energetic 3-nuclei with velocities below the electron
thermal velocities (but still above the lithium formation
reaction thresholds of $8.39\,$ and $7.05\,$MeV for \h3 and \he3,
respectively). I have computed the energy transfer of a fast charged
nucleon or nucleus due to Coulomb interactions with an electron-positron
plasma at temperature $T$, explicitly accounting for a thermal average
and accounting for higher order terms. Details on this calculation
will be presented elsewhere.
Since \li6 synthesis
during/after BBN is dominated during those epoch where (a) the annihilation
rate Eq.~(\ref{ann}) is still large, (b) synthesized \li6 survives the
thermal \li6$(p,\alpha )$\he3 reaction, and (c) ${\rm d}E/{\rm d}x|_C$
is at it's minimum, the bulk of the \li6 is synthesized at
$T\approx 10\,$keV and an exact evaluation of ${\rm d}E/{\rm d}x|_C$
is paramount to an evaluation of \li6 abundances.

\bef
\epsfxsize=8.5cm
\epsffile[85 50 410 302]{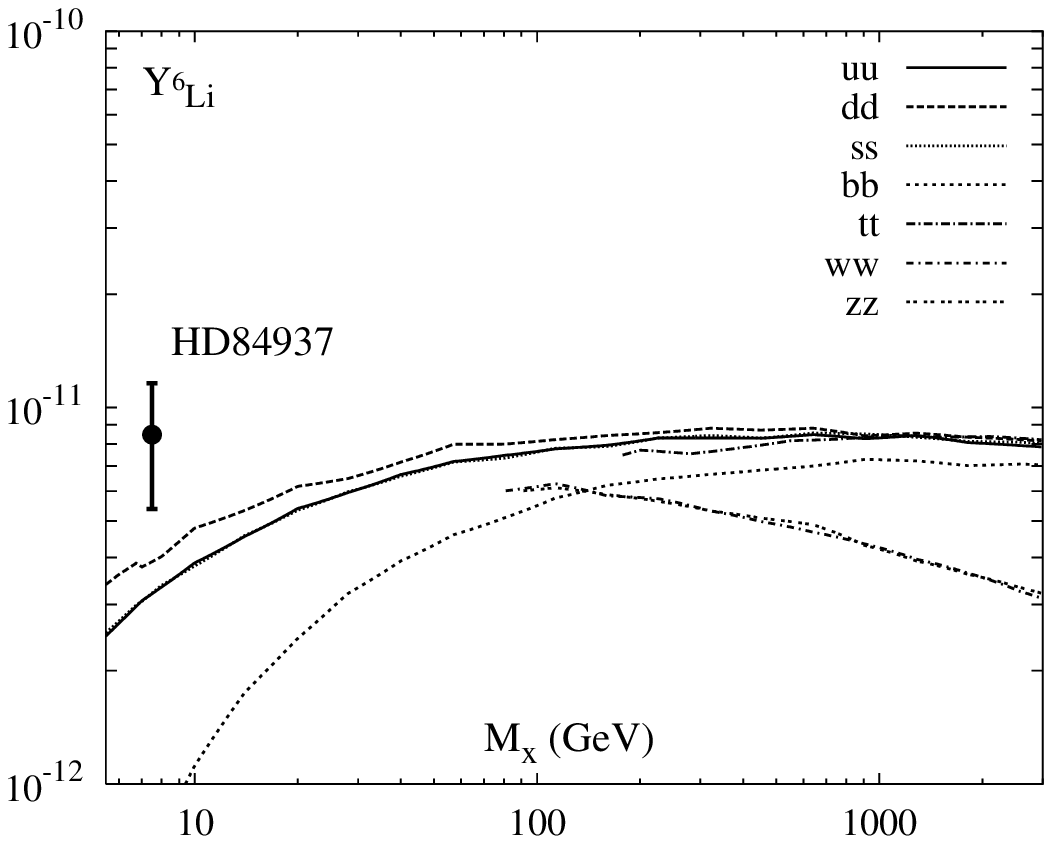}
\caption{Final \li6 yield functions defined via 
Eq.~(\ref{yield}) as a function of neutralino
mass for various annihilation channels as labeled in the key. The
1-$\sigma$ range of the \li6 abundance in HD84937 is also shown.}
\label{fig1}
\eef

\bef
\epsfxsize=8.5cm
\epsffile[85 50 410 302]{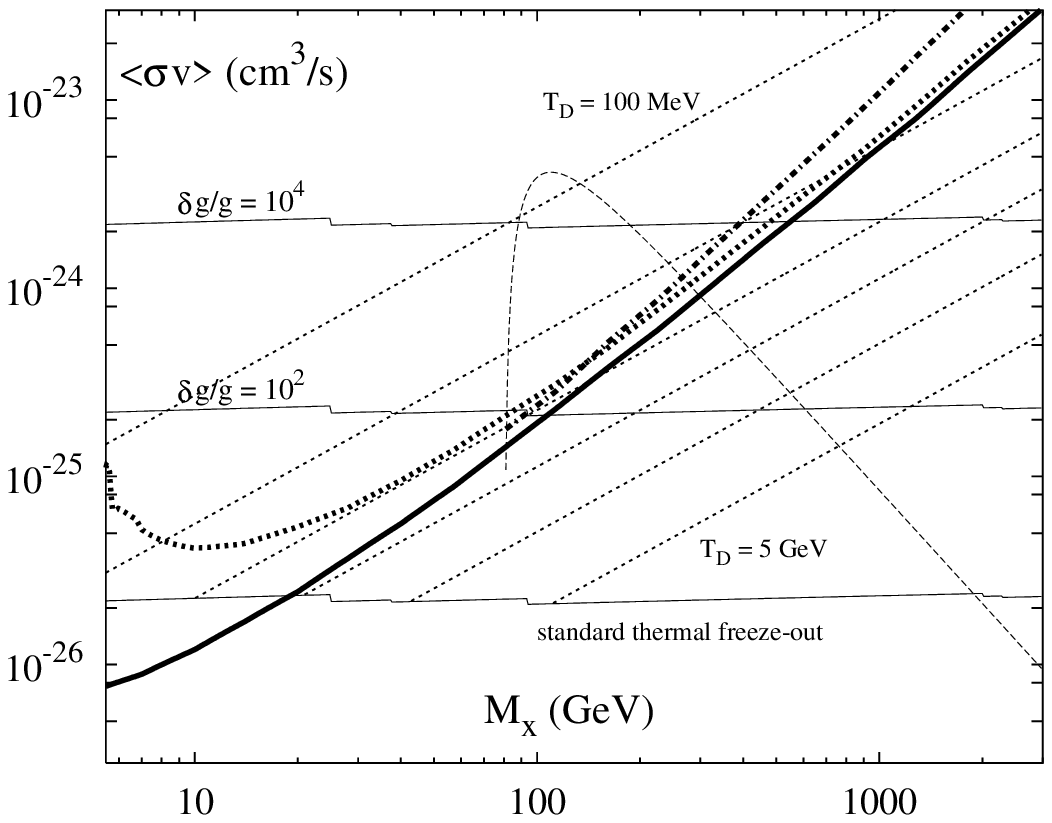}
\caption{Annihilation-channel dependent constraints on the
s-wave annihilation rate due to possible \li6 overproduction
as a function of neutralino mass. For simplicity, only the
$u\bar{u}$, $b\bar{b}$ and $W^-W^+$ channels are shown with
heavy lines and
line styles as indicated in Fig.1. Constraints on other 
quark- and gauge boson- annihilation channels are virtually identical to 
those shown as may be verified by inspection of Fig.1. Annihilation
rates above the lines are ruled out.
Also
shown are the annihilation rates required~\cite{remark4a} to produce 
$\Omega_{\chi}h^2 = 0.1126$ during standard thermal freeze-out (solid
line) or during thermal freeze-out when extra degrees of
freedom (contributing to the Hubble expansion during freeze-out) 
are present (the upper
two solid lines with $\delta g_H^f/g_H^f$ as labeled). The dotted
diagonal lines correspond to the required
$\langle\sigma v\rangle$ for post thermal freeze-out
non-thermal generation of 
$\Omega_{\chi}h^2 = 0.1126$ at temperatures, from top to bottom, $0.1,0.2,0.5,1,2,$
and $5\,$GeV, respectively. Here the QCD phase transition has been assumed to 
occur at $200\,$MeV. The curved dotted line shows the
annihilation rate in case of AMSB winos~\cite{Moroi:1999zb}.}
\label{fig3}
\eef

\bef
\epsfxsize=8.5cm
\epsffile[85 50 410 302]{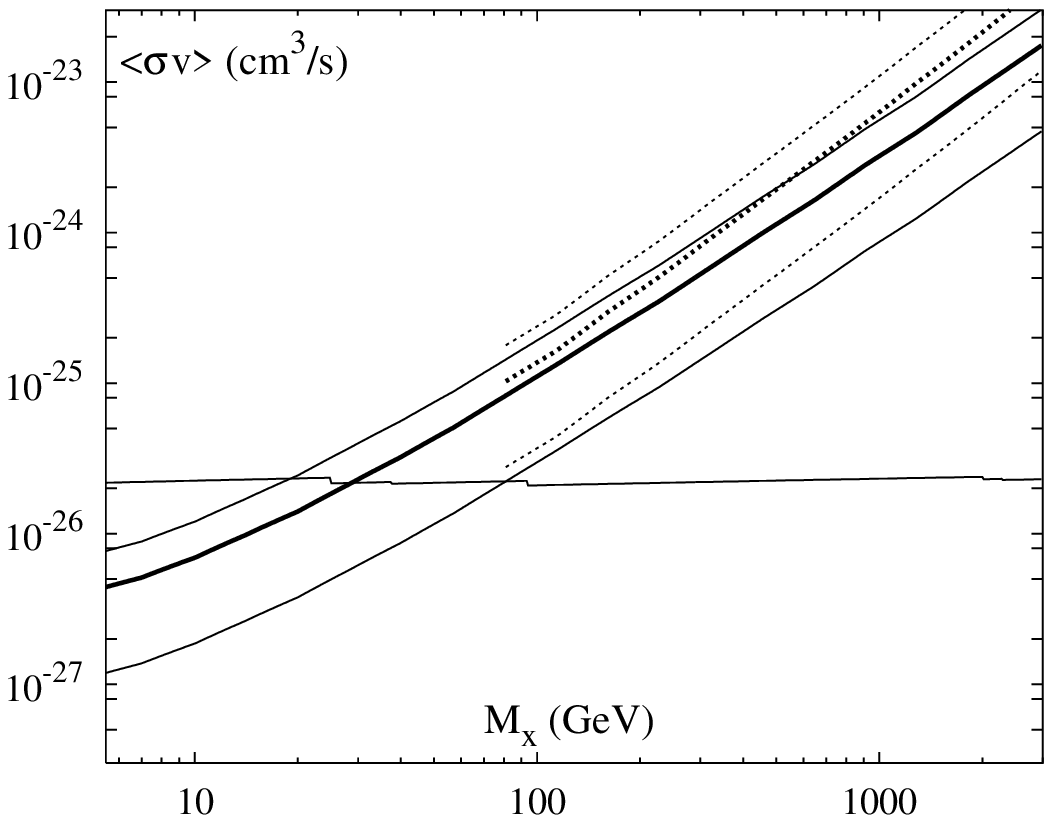}
\caption{S-wave annihilation rate required to produce within the
$2-\sigma$ limits the \li6 abundance of HD84937. The heavy lines
indicate the central value of HD84937, whereas lighter lines the
$2-\sigma$ ranges. For simplicity only the $u\bar{u}$ (solid)
and $W^-W^+$ (dotted) channels are shown with results for other
channels similar (cf. Fig. 1).}
\label{fig2}
\eef

I have thus computed the \li6 yield for annihilating particles and under
the assumption of specific annihilation channels such as into
$u\bar{u}$, $d\bar{d}$, $s\bar{s}$, $b\bar{b}$, or $t\bar{t}$
quark-antiquark pairs, as well as $W^-W^+$ or $ZZ$ gauge bosons~\cite{remark4}. 
The results
of these computations for varying neutralino masses $m_{\chi}$
are shown in Fig. 1. Here the annihilation-channel dependent
yield functions $Y_{\rm ^6Li}^i$ are defined via the equation
\begin{equation}
{n_{\rm ^6Li}\over n_{\rm H}} = 
\langle\sigma v\rangle_{-25}\, m_{\chi,100}^{-3/2}\,
\left(\frac{\Omega_{\chi}h^2}{0.1126}\right)^2
\sum_i b_i Y_{\rm ^6Li}^i
\label{yield}
\end{equation}
where the $b_i$ are branching ratios into channel $i$ and
$m_{\chi,100}$ is the neutralino mass in units of $100\,$GeV. 
Eq.~(\ref{yield}) is remarkable as it allows a quite general
evaluation of the final \li6 abundance, independent of
the nature of the annihilating particle, as well as applicable for
essentially all relevant $\langle\sigma v\rangle$ and $m_{\chi}$.
Note that the $Y_{\rm ^6Li}$'s for leptonic channels, which are not
shown, are essentially zero due to the absence of injected nucleons.
Annihilation into Higgs bosons, on the other hand, yield $Y_{\rm ^6Li}$
only somewhat smaller than those shown in Fig.1, since Higgs bosons
typically decay into heavy quarks or massive gauge bosons.
A simple scaling with $\langle\sigma v\rangle$ was
possible due to the linearity in the \li6 production and destruction
mechanisms. Note that though for fixed dark matter density
${\rm d\, Ann/d}t\sim m_{\chi}^{-2}$  
a scaling in Eq.~(\ref{yield}) with $m_{\chi}^{-1.5}$ 
has been adopted. This is due to higher mass $\chi$ producing more
energetic primary nucleons which, in turn, produce a larger number of
secondary $p$ and $n$, yielding a final 
$n_{\rm ^6Li}/n_{\rm H}\sim m_{\chi}^{-1.5}$. 

Fig. 1 also shows
the one-sigma range 
\li6/H$\approx 8.47\pm 3.10\times 10^{-12}$~\cite{Cayrel}
of the observationally best determined \li6/H-ratio in a 
low-metallicity star, i.e. in HD84937, which has been analyzed
by several groups~\cite{li6:lowZ,Cayrel}. 
\li6 detections have been currently
claimed in three low-metallicity 
$\rm [Z]\simle -2$~\cite{li6:lowZ,Cayrel,Niss:00}
and two higher metallicity stars 
$\rm [Z]\sim -0.6$~\cite{li6:highZ} 
with their abundances 
coincidentally all in the same range, reminiscent of a unique cosmic
abundance.
In principle, \li6 may be depleted in stars, in practice, however, 
those low $[Z]$ stars which show \li6 have normal ``Spite''-plateau \li7
abundances,
and as this latter isotope would, in most circumstances, be depleted as well,
substantial \li6 depletion seems unlikely~\cite{remark8}.  
The origin of the \li6 at $\rm [Z]\simle -2$
is somewhat mysterious as only with considerable
difficulty explained by traditional galactic cosmic ray spallation
and fusion reactions~\cite{li6_cosmicrays}. 
It is thus conceivable that \li6
at low $[Z]$ is, in fact, entirely of primordial origin, though other
alternative origins have also been proposed~\cite{Suzu:02}.

In the absence of stellar \li6 depletion the efficient production of \li6
due to annihilating neutralinos may be used to constrain the properties of
the dark matter particle itself. 
Fig. 2 shows annihilation-channel dependent limits on the s-wave 
annihilation $\chi\chi$ cross section of a dark matter particle, 
representing a convolution of
the results in Fig. 1 with Eq.~(\ref{yield}). Here the two-sigma upper
limit on \li6 in HD84937 has been adopted such that dark matter particles
above the diagonal lines in the upper part of the figure are
ruled out due to \li6 overproduction. For reference, the figure also
shows annihilation cross sections required to obtain the WMAP abundance
given (a) standard thermal freeze-out~\cite{remark4a} 
(b) thermal freeze-out
when additional degrees of freedom
are present $\delta g_{H}^f/g_{H}^f >0$ as, 
for example due to a quintessence field
and (c) via late generation below the thermal freeze-out temperature
due to, for example, decay of gravitinos
or evaporation of Q-balls. 
Furthermore, the figure shows
the predicted LSP wino $\langle\sigma v\rangle$
into $W^+W^-$ in AMSB scenarios as given in 
Ref.~\cite{Moroi:1999zb}

It is evident that possible \li6 overproduction imposes
stringent constraints on dark matter particle properties.  
As evident from Fig.2, light
neutralinos with mass $m_{\chi}\simle 20\,$GeV annihilating into light
quarks are ruled out at two-sigma. AMSB winos may only be 
consistent with the \li6 abundance when $m_{\chi}\simge 250\,$GeV.
When the dark matter particle is the first excited Kaluza-Klein
mode $B^{(1)}$~\cite{Servant:2002} of the $U(1)_Y$ gauge boson, it may
not be lighter than $m_{KK}\simle 200\,$GeV due
to its efficient annihilation into right-handed up-type quarks. 
Late-time generation of dark matter $\chi$'s is disallowed for
all temperatures below $200\,$MeV unless the dark matter particle
is fairly heavy $m_{\chi}\simge 400\,$GeV or annihilation does not
occur into hadronic or gauge (and higgs) boson channels. 
Finally, a substantially
increased Hubble rate during thermal freeze-out is for all but the
heaviest neutralinos observationally disallowed. These limits may be
circumvented in case substantial stellar \li6 depletion occurred which,
nevertheless, would require a coincidentally similar amount of
depletion in all three \li6-rich observed stars to date. 
Limits of this sort
are also important in light of scenarios which invoke neutralino
annihilation as putative explanations of, for example, the observed
cosmic positron excess at $\sim 10-30\,$GeV~\cite{Baltz:2001ir,Boer} 
as determined
by HEAT, or the 
galactic-center~\cite{Bertone:2002je,Cesarini:2003nr,Hooper:2004vp}, 
or extragalactic diffuse~\cite{Ullio:2002pj}, 
gamma-ray fluxes as determined by EGRET, VERITAS,
or CANGAROO. In order to explain anomalous components of such signals such
as bumps in the spectrum a signal boost (enhancement) factor $B_s$ of the order
$\sim 50-10^3$~\cite{Bergstrom:1998jj,Calcaneo-Roldan:2000yt} 
is essentially always required. Considering the most
recent N-body simulations on substructures and halo 
profiles~\cite{Stoehr:2003hf},
such $B_s$ seems only unlikely due to clumpy halos  
or singular halo profiles though, in principle, 
one could envision it due to an enhanced
$\langle\sigma v\rangle$ with respect to it's standard
thermal freeze-out value. Nevertheless, already modest particle-physics
motivated boost-factors of the order of $\sim 1-10$ will have to face a
potential \li6 overproduction problem.
 
Last but not least, it is possible that the entire observed \li6 at
low metallicity may be due to the residual annihilation of a dark
matter particle. Fig.3 shows the mass-dependent annihilation rate
required to produce a \li6 abundance within the 2$\sigma$ 
ranges of those observed in HD84937. It is seen that this may be accomplished
even by a standard thermal freeze-out with dominant s-wave
component annihilation into light quarks, provided the neutralino mass
is within the approximate range of $20-80\,$GeV~\cite{remark4a}. 
The observed amount
of \li6 may be produced for even larger mass neutralinos when either
coannihilation effects or annihilation on poles occur in the thermal
freeze-out case or neutralinos are generated non-thermally.
Coincidentally, the recently proposed specific dark matter 
neutralinos~\cite{Elsaesser:2004}
and Kaluza-Klein particles~\cite{Hooper:2004} 
which could explain the claimed bump in
the extra-galactic $\gamma$-ray background and/or the positron excess as
observed by HEAT would have just the right properties to yield \li6
abundances as observed in low-[Z] stars.
Such annihilation is
also associated with some, albeit small, amount of observationally 
favored \li7 depletion~\cite{Jeda:04}. 
It is intriguing that the
observed abundances of \li6
in low-metallicity stars may be entirely a product of 
dark matter annihilation.

\vskip 0.1in
I acknowledge discussions with Eric Nuss.


\begin{thebibliography}{99}
\bibitem{xtf} Here $x_{tf} = {\rm ln}(A)-(n+1/2){\rm ln(ln}(A))$
with $A\approx 0.038(n+1)M_{pl}M_{\chi}\sigma_0 g_{\chi}/(g_H^f)^{1/2}$
where $M_{pl}$ is the (non-reduced) Planck mass and $g_{\chi}$ is
the statistical weight of the freezing out particle.
\bibitem{remark7} Limits based on \h2 are typically more than
an order of magnitude weaker.
\bibitem{remark1} Note that though Eq.~(\ref{Prob}) incorrectly 
approximates mass three nuclei during p-\h3 (\he3) elastic scattering 
to loose most of their energy, in the Monte-Carlo analysis this
problem is treated properly.
\bibitem{remark3} Note, this finding is in stark contrast with the
result of the recent paper Ref.~\cite{Kawasaki:2004yh}, 
which makes me suspect that the authors 
of the latter analysis have not included a thermal average over electron
velocities in the computation of ${\rm d}E/{\rm d}x|_C$.
\bibitem{remark4} The calculations assume $\Omega_bh^2\approx 0.0218$.
Furthermore the small amount of standard
thermal \li6 production via D$(\alpha ,\gamma )$\li6~\cite{Noll:97} 
has been conservatively 
neglected. Annihilation products have
been forced in PYTHIA to be on-shell. The post-annihilation
``initial'' state is therefore two on-shell 
primaries with each energy $m_{\chi}$.
\bibitem{remark8} Note that results of analysis
of the observational data are typically given as \li6/\li7 ratios.
Conversion to a \li6/H-ratio requires thus a \li7/H ratio. This latter
value is given for HD84937 in Ref.~\cite{Cayrel} at
\li7/H$\approx 1.63\times 10^{-10}$, agrees with other
\li7/H ratios on the Spite-plateau~\cite{Ryan:99} but not with that
predicted from standard BBN for $\Omega_bh^2\approx 0.0218$, which is
about a factor $3$ larger. In case the observed value of \li7 is due
to a significant factor $\sim 3$ stellar depletion, a factor $\simge 3$
stellar depletion of \li6 is predicted. In this case, observational
constraints based on \li6 abundances are relaxed. On the
other hand, the resulting uncomfortably large pre-depletion
\li6 abundance seems unattainable within the context of
galactic cosmic ray nucleosynthesis, arguing for an origin of \li6 
which is likely primordial or at the least pre-galactic.
\end{thebibliography}
\end{document}